\documentclass[reprint,aps,pra,footinbib]{revtex4-1}
\usepackage{graphicx}
\usepackage{color}
\usepackage{pgfplots}
\usepackage{calc}
\usepackage{amsmath,amsfonts,amsthm,mathrsfs,bm,bbm}							 
\usepackage{braket}
\usepackage{bbold}
\usepackage{float}
\usepackage{fancyhdr}
\usepackage[citecolor=black,linkcolor=black]{hyperref}

\begin{document} 

\title{Application of quantum Darwinism to a structured environment}

\author{Graeme Pleasance}
\email{gpleasance1@gmail.com} 

\author{Barry M. Garraway}

\affiliation{Department of Physics and Astronomy, University of Sussex, Falmer, Brighton, BN1 9QH, United Kingdom}
\date{\today}

\begin{abstract}
Quantum Darwinism extends the traditional formalism of decoherence to explain the emergence of classicality in a quantum universe. A classical description emerges when the environment tends to redundantly acquire information about the pointer states of an open system. In light of recent interest, we apply the theoretical tools of the framework to a qubit coupled with many bosonic subenvironments. We examine the degree to which the same classical information is encoded across collections of: (i) complete subenvironments and (ii) residual ``pseudomode'' components of each sub-environment, the conception of which provides a dynamic representation of the reservoir memory. Overall, significant redundancy of information is found as a typical result of the decoherence process. However, by examining its decomposition in terms of classical and quantum correlations, we discover classical information to be non-redundant in both cases (i) and (ii). Moreover, with the full collection of pseudomodes, certain dynamical regimes realize opposite effects, where either the total classical or quantum correlations predominantly decay over time. Finally, when the dynamics are non-Markovian, we find that redundant information is suppressed in line with information backflow to the qubit. By quantifying redundancy, we concretely show it to act as a witness to non-Markovianity in the same way as the trace distance does for nondivisible dynamical maps. 
\end{abstract} 

\maketitle

\section{Introduction}\label{intro}

\begin{figure}[!t]
	\centering
	\includegraphics[width=.38\textwidth]{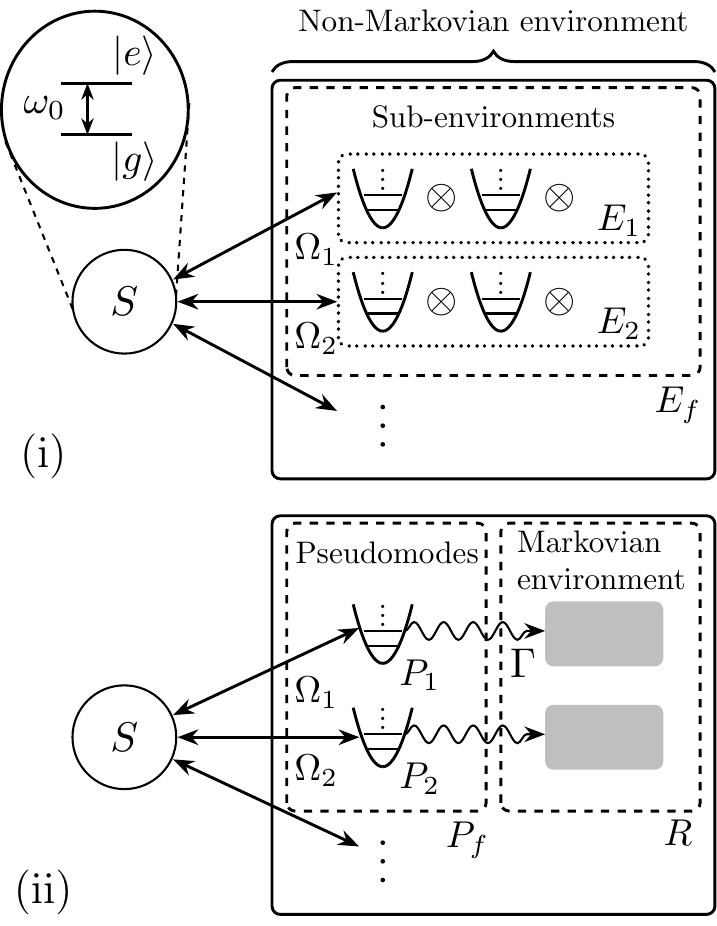}
	\caption{\label{Fig1}Schematic showing two dynamical representations of the model. (i) The original picture provided by Eqs. (\ref{eq:bare_Ham}) and (\ref{eq:Ham}), where the qubit system $S$ couples to many subenvironments $E_1,E_2,\dotso$ with strengths $\Omega_1,\Omega_2,\dotso$. (ii) An equivalent picture in terms of pseudomodes, labeled $P_1,P_2,\dotso$, which are each damped by independent Markovian reservoirs $R$ at a rate $\Gamma$. The environment is sampled by constructing fragments out of the bare subenvironments or the pseudomodes, indicated by $E_f$ and $P_f$, respectively.} 
\end{figure}

The theory of open quantum systems formulates the emergence of classical-like behavior in quantum objects through the action of environment induced decoherence \cite{Schlosshauer2004,ZurekQTC2004,BreuerTOQS2002,Zurek1991,JoosDecoh2013}. This is a phenomenon where superpositions are removed over time, leaving certain mixtures of stable macroscopic states, known as pointer states \cite{Zurek1981,Zurek1982,Zurek1993}. Inevitably, mitigating this process by, for example, utilizing reservoir engineering techniques \cite{Myatt2000,Verstraete2009} in an effort to successfully realize quantum technological devices relies on further understanding the fragility of quantum states and their information. \\
\indent During the decoherence process, the issue of how information about the pointer states is communicated to the external observer is usually forgone by ``tracing out'' and ignoring the active role of the environment. Quantum Darwinism instead proposes that a new step can be made \cite{Zurek2009}. When an open system $S$ interacts with its surroundings, correlations invariably develop between the two. Thus, to understand the nature of the information passed to the environment, it is useful to consider its role explicitly in the formalism rather than as a passive sink of coherence. In our paper we focus on some particular region of interest within the environment, labeled $X$, which we will contextualise shortly. Given that the system interacts with the degrees of freedom of $X$, the quantum mutual information (QMI) \cite{Nielson2010} can be used to ascertain what is known about the system by a fragment of this part of the environment, 
\begin{equation} \label{eq:QMI}
	I(\rho_{SX_f}) = S(\rho_S) + S(\rho_{X_f}) - S(\rho_{SX_f}),
\end{equation}
where $X_f$ makes up some fraction (in size) of $X$ and $S(\rho)=-\text{Tr}\left[\rho\,\text{ln}\,\rho\right]$ is the von Neumann entropy of $\rho$. The QMI defines a measure of correlations which act as a communication channel between the subsystems. \\
\indent The central thesis of quantum Darwinism is that the class of system-environment states produced by decoherence is unique, in that the states contain correlations encoding many local copies of classical data about the open system. This is done under a selection process during which only records specific to the pointer states can create lasting copies of themselves, and so end up passing their information into many different fragments. The redundancy $R_{\delta}$ \cite{Blume-Kohout2008, *Blume-Kohout2006, *Blume-Kohout2005, *Blume-KohoutPhDthesis} quantifies the average number of fragments that record up to classical information $I(\rho_{SX_f})=(1-\delta)S(\rho_S)$, with a small deficit $\delta$. To this effect, a large $R_{\delta}$ implies widely deposited classical data in the environment on the pointer observable\textemdash this being, in practice, the only knowledge of the quantum system accessible to measurements. \\
\indent In this paper we join this framework to a qubit system coupled to an ensemble of bosonic subenvironments. The motivation is two-fold. Firstly, it is of fundamental interest to understand whether the success of quantum Darwinism applies to an as of yet unexplored regime. It has in fact been shown that classical features are inherent to generic models of quantum Darwinism \cite{Brandao2015}, but it is unclear under what conditions these features manifest beyond the scope of this work. Secondly, we aim to consolidate recent investigations which found quantum Darwinism to be inhibited when memory effects arise in the dynamics \cite{Giorgi2015,Galve2016}. \\
\indent Our approach enacts a partitioning of the environment into its memory and non-memory parts, where $X$ is assigned to either one of two cases: (i) the full environment modes or (ii) the memory region (see Fig. \ref{Fig1}). The intent is to specifically look at where information is shared redundantly: also, in view of Refs. \cite{Zwolak2009,Zwolak2010}, we aim to examine how quantum Darwinism is affected for state (ii) which mixes over time by evolving through a noisy quantum channel. As we will show, in case (i) redundant information tends to always emerge at long times, which is also found in case (ii) though this can only be quantified within certain dynamical limits. However, judging whether true effects of quantum Darwinism appear from a large redundancy of the \textit{total} information does not give an accurate picture: writing the QMI in terms of the accessible information and the quantum discord shows most of the environment has to be measured to gain close to full classical data on the qubit. Further, the redundancy measure $R_{\delta}$ is used to explore how non-Markovian behavior of the open system affects its shared correlations with the environment. To gain a consistent interpretation of information backflow using Eq. (\ref{eq:QMI}), we check the time evolution of the redundancy against temporal changes in the trace distance for an arbitrary pair of input states to the channel. \\
\indent This paper is organised as follows. In Sec. \ref{Sec2} we introduce the dynamical solutions of the model via a master equation that describes the dynamics of the qubit and pseudomodes (memory). In Sec. \ref{Sec3} we address the application of the quantum Darwinism to our model. In Sec. \ref{Sec4} we compute the average of the QMI for different fragments and examine the multipartite structure of the total system-environment correlations, as quantified by Eq. (\ref{eq:QMI}), in terms of its classical and quantum components. Finally, we analyze the non-Markovian behavior of the model with regard to its affect on information redundancy. The paper is then summarized in Sec. \ref{Sec5}, while details of all methods are outlined in the Appendices.

\section{Dynamical Model}\label{Sec2}  

We start by considering a qubit $S$ interacting with an environment $E$ of bosons. The environment is arranged as $E=\otimes^{\#E}_{k=1}E_k$ and initially in a vacuum state, where the index $k=1,2,\dotso,\#E$ labels individual subenvironments $E_k$. Each are composed of harmonic oscillators with frequencies $\omega_{\lambda}$. The Hamiltonian of the qubit and environment are thus given by ($\hbar=1$)
\begin{equation}\label{eq:bare_Ham}
	H_S=\omega_0\sigma_+\sigma_-,\quad H_E=\sum_{k,\lambda}\omega_{\lambda}a^{\dagger}_{k,\lambda}a_{k,\lambda}.
\end{equation}
Here, the operator $\sigma_+$ ($\sigma_-$) raises (lowers) the qubit energy by an amount $\omega_0$, while $a_{k,\lambda}$ ($a^{\dagger}_{k,\lambda}$) is the bosonic annihilation (creation) operator of the $\lambda$-mode in the $k$ sub-environment. Note variables between subenvironments commute as $[a_{k,\lambda},a^{\dagger}_{k',\lambda'}]=\delta_{k,k'}\delta_{\lambda,\lambda'}$. Within the rotating wave approximation and interaction picture, the interaction Hamiltonian $H_I$ reads
\begin{equation}\label{eq:Ham}
	H_I(t) = \sum_k\left(\sigma_+\otimes B_k(t)+\text{h.c.}\right),
\end{equation}
where $B_k(t)=\sum_{\lambda}g_{k,\lambda}a_{k,\lambda}\text{exp}[-i(\omega_{\lambda}-\omega_0)t]$ and $g_{k,\lambda}$ is the coupling to the qubit. 

\subsection{Solutions to the model}

The pseudomode method \cite{Garraway1997,GarrawaySCR1997} provides complete solutions to the model by replacing the environment with a set of damped harmonic oscillators\textemdash the pseudomodes\textemdash which are identified through evaluating the poles of the spectral distribution $\rho_{\lambda}(g_{k,\lambda})^2$ when analytically continued to the complex plane. For a single shared excitation between the system and environment it is possible to derive a master equation describing the combined time evolution of the qubit and pseudomodes. Importantly, since the resulting equation is of Lindblad form, the original non-Markovian system is effectively mapped onto an enlarged qubit-plus-pseudomode system, the dynamics of which are Markovian. \\
\indent To retrieve the solutions using such an approach, it first proves useful to extract the frequency dependence of the system-environment coupling constants onto the structure function $D(\omega_{\lambda})$, 
\begin{equation}\label{eq:extract}
	\rho_{\lambda}(g_{k,\lambda})^2 = \frac{\Omega^2_k}{2\pi}D(\omega_{\lambda}),
\end{equation}
where $\rho_{\lambda}d\omega_{\lambda}$ gives the number of modes in the bandwidth $d\omega_{\lambda}$, and $\Omega_k$ is the coupling parameter of the qubit and $k$ sub-environment. The above is normalized according to
\begin{equation}\label{eq:norm2pi}
	\int^{\infty}_{-\infty}d\omega D(\omega) = 2\pi,
\end{equation} 
which introduces the total coupling strength $\Omega_0$ via the relation $\Omega^2_0=\sum_k\Omega^2_k=\sum_{k,\lambda}(g_{k,\lambda})^2$. In the following we consider structure functions of the form
\begin{equation}\label{eq:SF}
	D(\omega) = \sum_k\frac{w_k\Gamma}{(\xi_k-\omega)^2 + (\Gamma/2)^2},
\end{equation}
where $\Gamma$ is the width and $\xi_k$ is the peak frequency of an individual Lorentzian, each weighted by real positive constants $w_k$ satisfying $\sum_kw_k=1$. For simplicity, we have imposed that the Lorentzian widths\textemdash associated with the coupling of the system to each sub-environment\textemdash are equal. \\
\indent Suppose the qubit is initially prepared in the state $\ket{\psi}_S=c_g\ket{g}+c_e(0)\ket{e}$, with $\ket{g}$ ($\ket{e}$) its ground (excited) state. As the number of excitations are conserved in this model, the total state is restricted to the single excitation manifold, admitting the closed form 
\begin{equation}\label{eq:state} 
	\ket{\psi(t)} = c_g\ket{g,0} + c_e(t)\ket{e,0} + \sum_{k,\lambda}c_{k,\lambda}(t)\ket{g,1_{k,\lambda}},
\end{equation}
where $\ket{e,0}$ and $\ket{g,1_{k,\lambda}}$ indicate the excitation in the qubit and the $\lambda$-mode of the $k$ sub-environment, respectively. By substituting Eq. (\ref{eq:state}) into the Schr\"{o}dinger equation, we obtain a dynamical equation for $c_e(t)$ by eliminating the variables $c_{k,\lambda}(t)$. This leads to $\dot{c}_e(t) = -\int^t_0ds f(t-s)c_e(s)$, where the memory kernel $f(t-s)$ is obtained by taking the continuum limit over all subenvironments as follows: 
\begin{align}\label{eq:corrfun}
	f(t-s) &= \sum_{k,k'}\left[B_k(t),B^{\dagger}_{k'}(s)\right]\nonumber\\
		&=\frac{\Omega^2_0}{2\pi}\int^{\infty}_{-\infty}d\omega D(\omega)e^{-i(\omega-\omega_0)(t-s)}.
\end{align}
Because Eq. (\ref{eq:SF}) is meromorphic and contains simple poles in the lower half complex plane, we can rewrite the preceding equation of motion as
\begin{align} 
	\dot{c}_e(t) &= -i\sum_{k}\Omega_ke^{-i\Delta_kt}b_k(t),\label{eq:qubit} \\
	\dot{b}_k(t) &= -\frac{\Gamma}{2}b_k(t)-i\Omega_ke^{i\Delta_kt}c_e(t),\label{eq:pseudomode}
\end{align}
which are defined in a new rotating frame with respect to $\sum_k\Delta_ka^{\dagger}_ka_k$, and where $\Delta_k = \xi_k-\omega_0$. The coefficients 
\begin{equation} \label{eq:pseud}
	b_k(t)=-i\Omega_ke^{-\Gamma t/2}\int^t_0ds\,e^{(i\Delta_k-\Gamma/2)s}c_e(s)
\end{equation}
are interpreted as those of pseudomodes. We note the one-to-one correspondence between the number of poles contained in Eq. (\ref{eq:SF}) when extended to the complex $\omega$-plane and the resulting number of pseudomodes. \\
\indent The dynamics of the joint qubit-pseudomode degrees of freedom are formulated in terms of a Markovian master equation \cite{Lindblad1976,*Gorini1976}
\begin{equation} \label{eq:PMme}
	\dot{\rho}_{SP} = -i[H_0(t),\rho_{SP}] + \Gamma\sum_k\mathcal{D}[a_k]\rho_{SP},
\end{equation}
where $\mathcal{D}[a_k]\,\cdot \equiv a_k\cdot a^{\dagger}_k - \frac{1}{2}\{a^{\dagger}_ka_k,\cdot\}$ and $a_k$ ($a^{\dagger}_k$) is the annihilation (creation) operator of the $k$-pseudomode. This master equation is exact and describes the joint unitary time evolution of the qubit and pseudomodes, generated by the Hamiltonian
\begin{figure}[!t]
	\centering	
	\includegraphics[width=.47\textwidth]{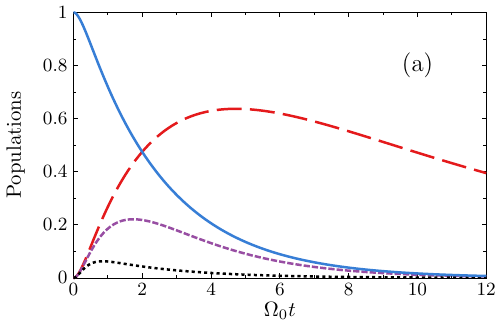}\\
	\includegraphics[width=.47\textwidth]{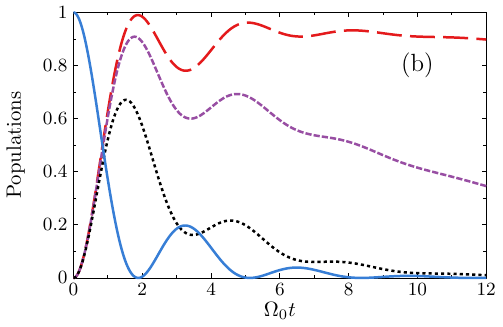} 
	\caption{\label{Fig2} Time evolution of the qubit Green's function (blue solid curve) from $|c_e(t)|^2=|G(t)|^2|c_e(0)|^2$ and excited pseudomode population for $c_e(0)=1$. Parameters are $\Gamma=\{10^{-2},0.1,0.5\}\Gamma_+$ (red long-dashed, violet dashed, black dotted curves) for $\Delta=0$. (a) Weak coupling case, $\Gamma_+=10\Omega_0$. (b) Strong-coupling (to moderate-coupling) case, $\Gamma_+=\Omega_0$.}
\end{figure}
\begin{equation}\label{eq:FH}
	H_0(t)=\sum_k\Omega_k\left(e^{-i\Delta_kt}\sigma_+\otimes a_k+\text{h.c.}\right).
\end{equation}
Hence, the original environment is equally represented in terms of a new structured one with a bipartite inner structure, composed of a set of uncoupled pseudomodes $P=\otimes^{\#E}_{k=1}P_k$ and Markovian reservoirs $R$. The qubit interacts directly with the pseudomodes, which each in turn leak into $R$ at a rate $\Gamma$ (see Fig. \ref{Fig1}). \\
\indent As Eqs. (\ref{eq:qubit}) and (\ref{eq:pseudomode}) are linear, exact solutions to the coefficients may be found directly through numerical inversion of the equations. However, we consider the continuum limit of pseudomodes, and, more generally, of the subenvironments by taking $\# E\rightarrow\infty$. This allows analytical solutions to be obtained provided a suitable distribution for the weights $w_k=w(\xi_k)$ is assumed in the conversion $\sum_{k}\rightarrow\int d\xi_k\rho_k$ (see appendix \ref{appenA}). A key result is that by choosing a single Lorentzian distribution of width $\Gamma_W$ Eq. (\ref{eq:SF}) can be written as
\begin{equation}\label{eq:SFplus}
	D(\omega) = \frac{\Gamma_+}{(\omega_0-\Delta-\omega)^2+(\Gamma_+/2)^2}. 
\end{equation}
The parameter $\Gamma_+=\Gamma+\Gamma_W$ is the resulting increased width compared to the single pseudomode case [where Eq. (\ref{eq:SF}) is a single Lorentzian of width $\Gamma$], while $\Delta$ is the detuning from the qubit frequency.

\subsection{Qubit and pseudomode dynamics}\label{2B}

The master equation (\ref{eq:PMme}) provides fully amendable solutions for strong system-environment interactions, which are shown in Fig. \ref{Fig2} to illustrate their behavior (matching solutions are in appendix \ref{appenA}). We initially check the response of the qubit by tracing out the pseudomodes from Eq. (\ref{eq:PMme}). This yields the time-convolutionless master equation $\dot{\rho}_S = \Gamma(t)[\sigma_-\rho_S,\sigma_+]+\Gamma^*(t)[\sigma_-,\rho_S\sigma_+]$ \cite{Anastopoulos2000}, where $\Gamma(t)=-\dot{G}(t)/G(t)$ and $G(t)$ is the Green's function of the qubit, i.e., $c_e(t)=G(t)c_e(0)$. By defining
\begin{align}
		\gamma(t)&= \Gamma(t)+\Gamma^*(t) = \sum_k2\Omega_k\frac{\text{Im}\left[e^{i\Delta_kt}c_e(t)b^*_k(t)\right]}{|c_e(t)|^2},\label{eq:timedepdecayrate}\\
		s(t)&=i\left[\Gamma^*(t)-\Gamma(t)\right] = \sum_k2\Omega_k\frac{\text{Re}\left[e^{i\Delta_kt}c_e(t)b^*_k(t)\right]}{|c_e(t)|^2},
\end{align}
the master equation of the qubit can be expressed in canonical form:
\begin{equation}\label{eq:SMe}
		\dot{\rho}_S=-i\frac{s(t)}{2}[\sigma_+\sigma_-,\rho_S]+\gamma(t)\mathcal{D}[\sigma_-]\rho_S,
\end{equation}   
where $\gamma(t)$ is a time-dependent decay rate and $s(t)$ is a Lamb shift. The dynamics associated with Eq. (\ref{eq:SMe}) are known to be nondivisible and non-Markovian if $\gamma(t)$ takes on negative values. It is instructive to consider this aspect with regard to the behavior of the qubit-pseudomode populations, as described by the following relation: 
\begin{equation}\label{eq:gm(t)}
	\sum_k\left(\frac{\partial}{\partial t}+\Gamma\right)|b_k(t)|^2=\gamma(t)|c_e(t)|^2.
\end{equation} 
Note this has been derived using Eqs. (\ref{eq:qubit}), (\ref{eq:pseudomode}) and (\ref{eq:timedepdecayrate}). The lefthand side of the above shows the rate of change the pseudomode population compensated against irreversible losses, which occur at a rate $\Gamma$. To examine this further, first we set $\Delta=0$, this being the case we shall focus on from now on. In the strong-coupling regime \footnote{Note that throughout we refer to the ``strong-coupling'' regime for parameters that yield oscillatory non-Markovian behavior in the qubit.}, defined by $4\Omega_0>\Gamma_+$, it is noticed that the qubit undergoes oscillatory dynamics\textemdash the excited population increases in time during intervals when $\gamma(t)<0$, which, from Eq. (\ref{eq:gm(t)}), gives a simultaneous and equal decrease in the pseudomode population, while, in the weak-coupling regime $4\Omega_0<\Gamma_+$, the time evolution of the qubit tends towards exponential decay at the Markov emission rate $\gamma_0=4\Omega^2_0/\Gamma_+$. In this instance $\gamma(t)$ is positive at all times and the qubit-pseudomode populations show no oscillations. We therefore interpret the non-Markovian behavior as being causal to the backflow of population and energy between the two. As discussed in Ref. \cite{Mazzola2009}, this indicates that the pseudomode region $P$ in Fig. \ref{Fig1} acts as a memory for the qubit in the presence of strong interactions. \\
\indent At this point it is also worth elaborating on the pseudomode population dynamics, obtained via the density matrix $\rho_{P}=\text{Tr}_S[\rho_{SP}]$. Before we go into this, we first highlight the fact that the time evolution of $|G(t)|^2$ depends only on the memory kernel (\ref{eq:corrfun}). In turn, this means the qubit dynamics is solely determined by the structure function (\ref{eq:SF}). One might intuitively expect something similar for the dynamics of the pseudomode coefficients. However, we actually discover two damping time scales that affect $b_k(t)$. Its solution, provided in appendix \ref{appenA}, separates into two parts, each with different exponential prefactors, causing one part containing functional terms to decay at a rate $\Gamma_+/4$, and another part with static terms to decay at a rate $\Gamma/2$. Because of ``mixing'' between terms in the population $\sum_k|b_k(t)|^2$, it is difficult to single out their individual effect in a typical time evolution, which generally shows complex behavior. It becomes apparent, though, when we introduce a large separation of time scales through
\begin{equation}\label{eq:largesep}
	\frac{1}{\Gamma_+}\ll t\ll\frac{1}{\Gamma}. 
\end{equation}
In Fig. \ref{Fig2}, the effect of the fast and slow terms becomes increasingly noticeable towards the regime $\Gamma\ll\Gamma_+$. We see the fast terms decay quickly and predominantly influence the short-time evolution, while the slow terms decline exponentially and thus survive into the long-time limit.  \\
\indent Let us now comment further on the dynamics in such a case where Eq. (\ref{eq:largesep}) is valid. Within the strong-coupling regime, there is a distinct cross-over owing to the fact that the fast oscillatory terms decay on the fixed timescale $t\sim O(1/\Gamma_+)$. The dynamics are then categorised into two phases. As we have seen, the short-time evolution is characterized by memory effects where the qubit and pseudomode populations oscillate in time. When $t\gg1/\Gamma_+$, the pseudomode population instead decays monotonically as [$c_e(0)=1$]
\begin{equation}\label{eq:quasibound}
	\sum_k|b_k(t)|^2 \approx e^{-\Gamma t}\sum_k\frac{16\Omega^2_k\left[\Delta^2_k+(\Gamma_W/2)^2\right]}{\left|\left(2i\Delta_k+\frac{1}{2}(\Gamma-\Gamma_W)\right)^2+\Omega^2\right|^2}
\end{equation} 
where $\Omega=\sqrt{4\Omega^2_0-(\Gamma_+/2)^2}$ [see Eq. (\ref{eq:b_k}) in appendix \ref{appenA}]. At this point the qubit has essentially relaxed and thus decoupled from the memory, i.e., $|G(\infty)|^2\approx 0$. The density matrix $\rho_P$ then obeys the master equation
\begin{figure*}[!t] 
	\centering
	\includegraphics[width=.47\textwidth]{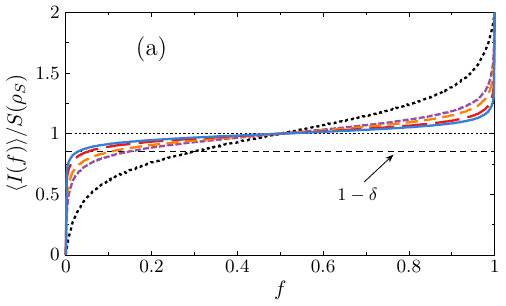} 
	\hspace{-.2cm}
	\includegraphics[width=.47\textwidth]{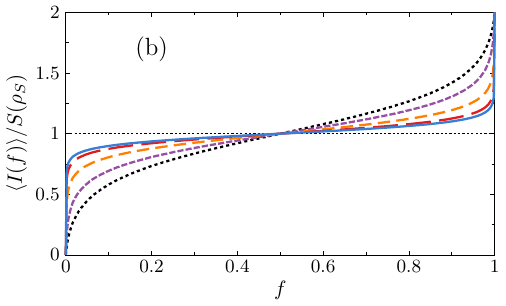}
	\caption{\label{PIP(i)} Partial information $\langle I(f)\rangle/S(\rho_S)$ between the qubit and subenvironments shown as a function of fraction size $f$ at times $\Omega_0t=\{10,15,25,40,50\}$ (black dotted, violet short-dashed, orange dashed, red long-dashed, and blue solid curves), for parameters (a) $\Gamma_+=\Omega_0$ and (b) $\Gamma_+=10\Omega_0$. In the shown examples, the spreading of correlations over time leads to emergent redundant information, indicated by the presence of the flat plateau. The intercept between the black dashed line (shown for $\delta=0.15$) and each curve highlights the average size of a fraction $f_{\delta}=1/R_{\delta}$ [Eq. (\ref{eq:red})] containing up to $(1-\delta)S(\rho_S)$ information on the qubit pointer states.}
\end{figure*}
\begin{equation}\label{eq:quasiboundME}
	\dot{\rho}_P = \Gamma\sum_k\mathcal{D}[a_k]\rho_P, \quad t\gg1/\Gamma_+. 
\end{equation}
Similar phases also exist in the pseudomode dynamics when the qubit undergoes a Markovian evolution. Notice here, however, that the cross-over is not as distinct since the fast terms do not decay on the same fixed time scale. Nevertheless, there is still a transition to slow exponential decay close to when the qubit has fully dissipated its energy. \\
\indent When $t\ll1/\Gamma$, Eq. (\ref{eq:quasibound}) predicts that the pseudomodes tend to form a quasi-bound state at long times $\Omega_0t\gg 1$ as a result of the cross-over, i.e., $\Gamma\ll\Gamma_+$. Although this occurs generally with respect to the coupling $\Omega_0$, the excitation is most efficiently ``trapped'' by the pseudomodes in the strong-coupling limit since increasing $\Gamma_+$ (with the ratio $\Gamma/\Gamma_+$ fixed) also increases the rate at which population leaks to the Markovian reservoir. Overall, we find the validity of Eq. (\ref{eq:quasibound}) in describing the long-time dynamics to only really be affected by the degree of separation between $\Gamma$ and $\Gamma_+$. The trapping effect then appears to be a feature of the narrow-Lorentzian structure of $D(\omega)$ [Eq. (\ref{eq:SF})]. Indeed, by taking the broad Lorentzian limit $\Gamma\approx\Gamma_+$ one recovers the usual single pseudomode dynamics from Ref. \cite{GarrawaySCR1997}, which does not display any of the trapping features seen here. \\
\indent The presence of a large pseudomode population well into the long-time limit suggests that a significant proportion of the total correlations of $S+E$ develops between the qubit and memory region of the environment. Since we are working within the context of quantum Darwinism, it seems justified to ask if such correlations translate into redundant information. This is part of what we consider in forthcoming sections.

\section{Application of Quantum Darwinism}\label{Sec3}  

In this section we move onto investigate emergent features of quantum Darwinism. Here, the central quantity under study is the QMI (\ref{eq:QMI}), which is computed for a given choice of fragment. As we are working in a typical decoherence setting \cite{Zurek1991} the most obvious choice is to construct fragments out of the subenvironments. In this paper we would also like to go further to address \textit{where} information regarding the pointer states manifests within the environment. A natural way to go about this is to additionally construct fragments out of the memory part of the environment, using the pseudomodes. \\
\indent Let us recall $X$, which was originally used to indicate some region of interest in the environment [see Eq. (\ref{eq:QMI})]. In view of the bipartite structure of the environment, we will consider two cases where a fragment $X_f=\otimes^{\{m\}}_{k=1}X_k$ is made up using a random combination $\{m\}=\{k_1,\dotso,k_m\}$ of either (i): the bare subenvironments $E_k$, or (ii): the pseudomodes $P_k$. Both schemes are depicted in Fig. \ref{Fig1}. We emphasise that the idea of using a two-pronged approach to check the locality of redundant information is entirely based on the fact that the two representations of the environment coincide. \\
\indent Successful quantum Darwinism is characterized by a large redundancy, $R_{\delta}\gg1$, needing the full mutually induced decoherence of many system-fragment combinations for the environment to learn anything about the qubit. Our objective is to then use the QMI to gauge the extent to which system-environment correlations produced under the time evolution of the pure state (\ref{eq:state}), and/or the density matrix $\rho_{SP}$ (\ref{eq:PMme}), are universally shared between fragments: or, more simply put, how many fragments communicate roughly the same information about the qubit, in each case. To this end, we quantify the size of a given fragment using
\begin{equation}\label{eq:f}
	f = \frac{m}{\# E},
\end{equation}
where $f$ is the fraction, $0\leq f\leq 1$, and $m=1,2,\dotso, \#E$ is the number of objects in the fragment: e.g., for $f=1$, the full environment is sampled and so $X_{f=1}=X$. The computation of the redundancy follows directly from the ``partial information'' $\langle I(\rho_{SX_f})\rangle=\langle I(f)\rangle$ \cite{Zurek2009}, where $\langle\cdot\rangle$ denotes the average over all fragments of size $f$. In terms of Eq. (\ref{eq:f}), the redundancy is given by 
\begin{equation}\label{eq:red}
	R_{\delta} = \frac{1}{f_{\delta}},
\end{equation}
in which 
\begin{equation}\label{eq:classplat}
	\langle I(f_{\delta})\rangle =(1-\delta)S(\rho_S).
\end{equation}

\section{Results}\label{Sec4}

In order to examine the redundant recording of information in cases (i) and (ii), we employ a Monte Carlo procedure to randomly sample fragments of size $f$ for every $m=1,2,\dotso, \#E$. The partial information is then computed by averaging  the QMI over the ensemble. 

\subsection{Quantum Darwinism}\label{Sec4a}

Numerical results are obtained for both Markovian and non-Markovian dynamics of the qubit, assuming an initial system-environment state $\rho_{SE}(0)=\ket{e}\bra{e}\otimes\ket{0}\bra{0}$. We proceed by discussing each of the cases (i) and (ii) in turn. 
\subsubsection{Case (i)}
Figure \ref{PIP(i)} shows the partial information plots of the qubit and subenvironments. At short times, the qubit quickly develops correlations with the environment until its state is maximally mixed, from which point the average entropy of the system decreases as it relaxes to the ground state. For a pure global state, as in Eq. (\ref{eq:state}), the plots are antisymmetric about $f=1/2$: mathematically, the sum of the (average) QMI between the qubit and any two complimentary fragments $E_f$ and $E_{1-f}$ of the environment satisfies \cite{Blume-Kohout2008, *Blume-Kohout2006, *Blume-Kohout2005, *Blume-KohoutPhDthesis}
\begin{equation}\label{eq:antisym}
	I(\rho_{SE_f})+I(\rho_{SE_{1-f}})=2S(\rho_S),
\end{equation}
which stems from the fact that the marginal entropies of $SE_{f}\otimes E_{1-f}$ are equal, i.e., $S(\rho_{SE_f})=S(\rho_{E_{1-f}})$. The bipartite $S+E$ also means the sum of the information from complimentary fragments provides the total information $I(\rho_{SE})=2S(\rho_S)$. As decoherence sets in, we see that the partial information increases more quickly to the maximum value around $f\approx 1$, in turn matched by a gradually steeper gradient close to the origin. This channels the middle region of the plot into a flat plateau shape, where its length indicates the availability of classical information $(1-\delta)S(\rho_S)$ from fractions of the environment. The redundancy typically measures the length of this plateau. Snapshots are displayed at increasing times where the plateau begins to level out, indicating that eventually many fractions gain access to the same information on the qubit for both the strong (moderate) and weak system-environment coupling. \\
\indent However, when the dynamics is strongly non-Markovian, i.e. for $\Omega_0\gg\Gamma_+$, we find the partial information oscillates about $f=1/2$ and hence no stable plateau develops (not shown). The origin of such behavior is exposed by examining the dynamics of $S(\rho_{E_f})$ for a typical fragment, which for weak coupling, grows monotonically until $S(\rho_{E_f})\approx S(\rho_{SE_f})$, so that $I(\rho_{SE_f})\approx S(\rho_S)$. If we increase $\Omega_0$ enough, the same entropy oscillates over time and for very strong coupling these oscillations continue in the limit $t\rightarrow\infty$. This disrupts the process by which the fragment acquires information due to periodic intervals of decorrelation. We see more precisely how the exchange of population between the qubit and memory affects the plateau in Sec. \ref{Sec4d}.  \\
\indent Before moving on to the next case we examine a reduction of Eq. (\ref{eq:QMI}) to a much simpler analytical form, which we can use to check the accuracy of our numerical results. This is first achieved by mapping the density matrix of a fragment state onto a single qubit \cite{Lazarou2012}. Note the mapping is not specific to either case (i) or (ii), and, accordingly, we shall use it to approximate the QMI in both such cases. The ground state of the collective qubit is universally defined as $\ket{\widetilde{0}}_{X_f}=\ket{\{0\}}_{X_f}$, while here ($X_f=E_f$) the excited state is formed using
\begin{equation}\label{eq:col_qubit_E}
	\ket{\widetilde{1}}_{E_f} = \frac{1}{\eta_{E_f}(t)}\sum_{k\in E_f,\lambda}c_{k,\lambda}(t)\ket{1_{k,\lambda}},
\end{equation}
where $k\in X_f$ denotes summation over objects in the fragment, and
\begin{equation}
	\eta_{E_f}(t) = \sqrt{\sum_{k\in E_f,\lambda}\left|c_{k,\lambda}(t)\right|^2}.
\end{equation}
By approximating
\begin{equation}\label{eq:eta_E}
	\eta^2_{E_f}(t)\approx f\eta^2_E(t) =f\sum_{k,\lambda}\left|c_{k,\lambda}(t)\right|^2,
\end{equation}
for all fraction sizes, the joint system-fragment state can be written as
\begin{equation}\label{eq:rho_SEf}
	\rho_{SE_f} = 
	\begin{pmatrix}
		(1-f)\eta^2_E & 0 & 0 & 0\\
		0 & f\eta^2_E &\sqrt{f}c^*_e\eta_E & 0\\
		0 & \sqrt{f}c_e\eta_E & |c_e|^2 & 0\\
		0 & 0 & 0 & 0
	\end{pmatrix}
	,
\end{equation}
being taken in the basis $\{\ket{g,\widetilde{0}_{E_f}},\ket{g,\widetilde{1}_{E_f}},\ket{e,\widetilde{0}_{E_f}},\ket{e,\widetilde{1}_{E_f}}\}$. The eigenvalues of $\rho_{SE_f}$ provide the following expression for the partial information of the qubit and fragment state,
\begin{align}\label{eq:QMI_2qE}
	I(\rho_{SE_f}) &= h\left(|c_e(t)|^2\right)+h\left(\chi_E(f)\right)-h\left(\chi_E(1-f)\right),
\end{align}
where 
\begin{equation}
	\chi_E(f)=f\eta^2_E(t), 
\end{equation}
and $h(x) = -x\,\text{ln}\,x - (1-x)\text{ln}(1-x)$. In Figs. \ref{Fig_a_pred}(a-b) we see Eq. (\ref{eq:QMI_2qE}) reproduces the numerical results remarkably well, with only small discrepancies appearing at limiting values of the fraction size for $\Gamma_+=\Omega_0$, close to the boundaries of the plots at $f=0$ and 1. Thus, our simple analytical model manages to predict the key features of the partial information plots to a good degree of accuracy.  

\subsubsection{Case (ii)}

The global state becomes mixed beyond $t=0$ and so the partial information plots do not acquire the same form. The QMI instead satisfies the inequality
\begin{equation}\label{eq:Infobound}
	I(\rho_{SP_f})\leq I(\rho_{SE_f}),\quad t\geq 0,\,\forall f,  
\end{equation}
where the upper bound is set by the strong sub-additivity of the von Neumann entropy \cite{Nielson2010}. As the equality only strictly holds at $t=0$, Eq. (\ref{eq:Infobound}) is understood from the idea that the state $\rho_{SP}$ evolves under a noisy quantum channel where information irreversibly leaks out to the Markovian reservoir. 
\begin{figure}[!t]
	\centering
	\includegraphics[width=.48\textwidth]{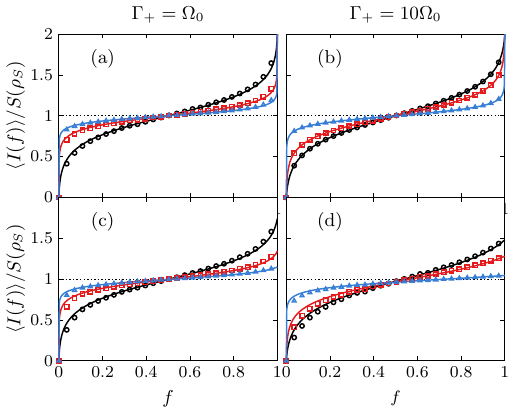}
	\caption{\label{a_pred} \label{Fig_a_pred} Analytical approximations of the partial information (solid curves) plotted against the corresponding numerical results at times $\Omega_0t=\{10,15,40\}$ (black circles, red squares and blue triangles), for $\Delta=0$. (a), (b) Average of $I(\rho_{SE_f})$ [Eq. (\ref{eq:QMI_2qE})]. (c), (d) Average of $I(\rho_{SP_f})$ [Eq. (\ref{eq:QMI_2qP})] for $\Gamma=10^{-3}\Gamma_+$. Note the left-hand column is for parameters $\Gamma_+=\Omega_0$, and the right-hand column is for $\Gamma_+=10\Omega_0$. It can be seen that the analytical results fit the numerics accurately within the plateau region at longer times.}
\end{figure}
The rate at which the vacuum state population increases signifies the noisiness of the channel. In view of this aspect, the top row of Fig. \ref{PIP(ii)} shows the partial information plots of the qubit and pseudomodes within the lossy regime $\Gamma_+\approx\Gamma$, where the vacuum population increases significantly at short times. Here, correlations between the qubit and pseudomodes typically decay quickly, though with strong system-environment interactions the QMI dissipates more slowly and redundant correlations have time to develop. In this instance we notice the appearance of a similar plateau feature from before. \\
\indent If we were to find the redundancy, however, it generally proves troublesome to compute using a fixed fraction size $f_{\delta}$  due to lack of antisymmetry of the partial information\textemdash that is, the plateau drops below the threshold $(1-\delta)S(\rho_S)$ for $\delta\ll 1$. This issue raises the question: are there circumstances where the application of the redundancy measure $R_{\delta}$  is possible? To answer this, we briefly look at how the time-dependent behavior of the purity $P(t)=\text{Tr}[\rho^2_{SP}(t)]$ changes with respect to the parameters $\Gamma$, $\Gamma_+$, and $\Omega_0$. Figure \ref{figPM} depicts $P(t)$ for different values of the spectral widths and coupling strength. First, when $\Gamma\ll\Omega_0$, we notice the purity decays to its minimum value on the time scale $t\sim O(1/\Gamma)$ from the fact that the gradient of $P(t)$ is approximately ten times larger between $\Gamma_+=\Omega_0$ and $10\Omega_0$ when $\Gamma=10^{-3}\Gamma_+$. For $\Gamma\ll\Gamma_+$, we can then expect the information content of $\rho_{SP}$ to stay closer to the equality of Eq. (\ref{eq:Infobound}) than the examples seen in Figs. \ref{PIP(ii)}(a) and (b) for $f=1$. This is because the purity declines more slowly when there is a large separation of time scales (assuming the same values of the coupling $\Omega_0$ are used from before). As such, the partial information plots should retain some of the antisymmetry features expressed by Eq. (\ref{eq:antisym}). \\
\begin{figure*}[!t] 
	\centering
	\includegraphics[width=.48\textwidth]{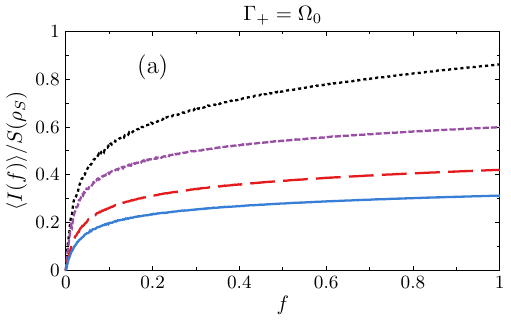}
	\hspace{-.2cm} 
	\includegraphics[width=.48\textwidth]{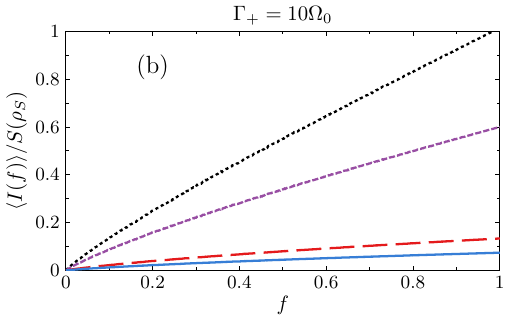}\\
	\vspace{-.2cm}
	\includegraphics[width=.48\textwidth]{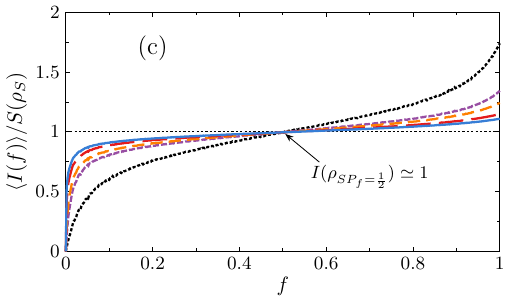}
	\hspace{-.2cm} 
	\includegraphics[width=.48\textwidth]{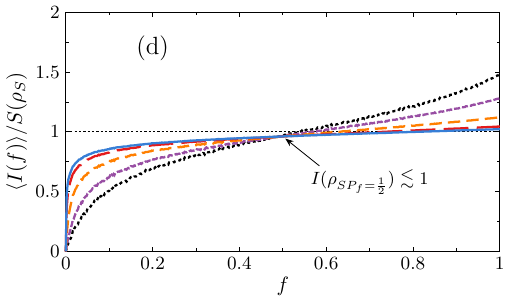}
	\caption{\label{PIP(ii)} Partial information $\langle I(f)\rangle/S(\rho_S)$ between the qubit and pseudomodes shown as a function of fraction size $f$ at times (a) $\Omega_0t=\{5,15,25,40\}$, (b) $\Omega_0t=\{0.5,1,5,10\}$ and (c),(d) $\Omega_0t=\{10,15,25,40,50\}$ ($t$ increases in the same order as each curve in Fig. \ref{PIP(i)} from dotted to solid). The left-hand column is for strong (moderate) coupling $\Gamma_+=\Omega_0$ while the right-hand column is for weak coupling $\Gamma_+=10\Omega_0$ (all $\Delta=0$). Unlike case (i), total correlations are erased over time as the state $\rho_{SP_f}$ evolves through a noisy quantum channel. (a), (b) $\Gamma=0.4\Gamma_+$: partial information decays quickly though for strong coupling redundancy features are qualitatively noticeable. (c), (d) $\Gamma=10^{-3}\Gamma_+$, a classical plateau is present in the long-time limit since the partial information retains approximate asymmetry about $f=1/2$ (indicated by the arrows), except at the boundary $f=1$.}
\end{figure*}
\indent Figures \ref{PIP(ii)}(c) and \ref{PIP(ii)}(d) show the development of a flat classical plateau over time for parameters $\Gamma\ll\Gamma_+$. Small differences in the partial information plots are apparent between the strong- and weak-coupling limits as the rate at which purity decays slightly increases with higher values of $\Gamma_+$ [see Fig. \ref{figPM}]. While Figs. \ref{PIP(ii)}(c) and \ref{PIP(ii)}(d) tend to deviate from a complete antisymmetric form at longer times, they maintain a similar shape to those in Fig. \ref{PIP(i)} for almost all fraction sizes below $f=1$. Crucially then, because the partial information saturates to the limit in Eq. (\ref{eq:classplat}) ($\delta\ll 1$) once there is sufficient decoherence of the state, the redundancy measure $R_{\delta}$ can be used even without the qubit-pseudomode state being pure. This is clear from comparing these plots between the two cases (i) and (ii) at equal times, where both exhibit a similar plateau. \\
\indent We can map the state of a fragment $P_f$ to that of a collective qubit, the excited state of which is defined by
\begin{equation}\label{eq:col_qubit_P}
	\ket{\widetilde{1}}_{P_f} = \frac{1}{\eta_{P_f}(t)}\sum_{k\in P_f}b_k(t)\ket{1_k},
\end{equation}
with normalization
\begin{equation}\label{eq:col_qubit_P}
	\eta_{P_f}(t) = \sqrt{\sum_{k\in P_f}\left|b_k(t)\right|^2}.
\end{equation}
Here, we look to follow a similar method that lead to a simple analytical expression for the partial information [see Eqs. (\ref{eq:eta_E}) and (\ref{eq:rho_SEf})], provided in Eq. (\ref{eq:QMI_2qE}), with the purpose of reproducing the results shown in Fig. \ref{PIP(ii)}.  If we again assume on average that
\begin{equation}
	\eta^2_{P_f}(t)\approx f\eta^2_P(t) = f\sum_k\left|b_{k}(t)\right|^2,
\end{equation}
then the density matrix $\rho_{SP_f}$ is given by
\begin{equation}
	\rho_{SP_f} = 
	\begin{pmatrix}
		\Pi_p+(1-f)\eta^2_P & 0 & 0 & 0\\
		0 & f\eta^2_P &\sqrt{f}c^*_e\eta_P & 0\\
		0 & \sqrt{f}c_e\eta_P & |c_e|^2 & 0\\
		0 & 0 & 0 & 0
	\end{pmatrix}
	,
\end{equation}
using the basis states $\{\ket{g,\widetilde{0}_{P_f}},\ket{g,\widetilde{1}_{P_f}},\ket{e,\widetilde{0}_{P_f}},\ket{e,\widetilde{1}_{P_f}}\}$. It turns out the partial information is then
\begin{equation}\label{eq:QMI_2qP}
	I(\rho_{SP_f}) = h\left(|c_e(t)|^2\right) + h\left(\chi^1_P(f)\right) - h\left(\chi^2_P(1-f)\right), 
\end{equation}
where the coefficients are 
\begin{align}
	\chi^1_P(f) &= f\eta^2_P(t),\nonumber\\
	\chi^2_P(f) &= f\eta^2_P(t)+\Pi_p(t),
\end{align}
and $\Pi_p(t)$ is the vacuum population of the pseudomodes (see definition in Appendix \ref{appenA}). In Figs. \ref{Fig_a_pred}(c) and \ref{Fig_a_pred}(d), results obtained from the approximate form of the partial information (\ref{eq:QMI_2qP}) are presented against the previously discussed numerical results at various times, with $\Gamma\ll\Gamma_+$ and $\Gamma_+=\Omega_0$. Our analytical formula shows remarkable agreement with the numerics, though small differences are noticeable: in particular, the partial information is slightly overestimated for small values of $f$. Regardless, the main features of these plots are captured, the most important being the increasing flattening of the plateau over time and subsequent emergence of redundant information. \\
\begin{figure}[!h]
	\centering
	\includegraphics[width=.46\textwidth]{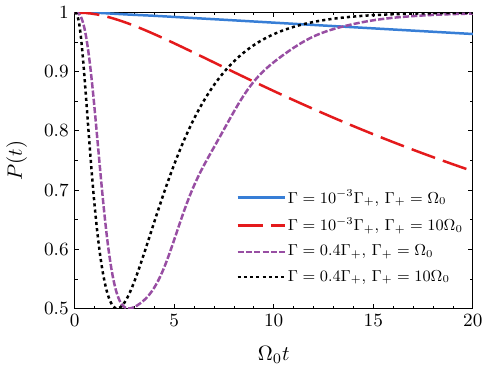}
	\caption{\label{figPM} Time evolution of the purity $P(t)=\text{Tr}[\rho_{SP}]$ shown for parameters $\Delta=0$, $\Gamma_+=\Omega_0$ (blue solid, violet dashed curves), and $\Gamma_+=10\Omega_0$ (red long-dashed, black dotted curves).}
\end{figure}
\indent Just as in case (i), a large redundancy in this case indicates widely accessible classical information. However, because this information is located in the pseudomodes it reveals more about the interaction: specifically, that records containing up to classical information on the qubit are held redundantly within the memory region of the environment. This differs from lossy interactions ($\Gamma\approx\Gamma_+$), where the damping noise (coming from the increase in vacuum population) severely restricts the time window in which redundant information can form before being lost completely (see Fig. \ref{PIP(ii)}). \\ 
\indent On the same point, it is noteworthy that the QMI of the full fraction of pseudomodes decays on a much faster time scale than the redundant information (in the plateau region). For a small damping rate $\Gamma$, it is reasonable to question if this corresponds to a loss of quantum information from $S+P$, since the plateau sits approximately at the classical limit with most information lost from global correlations. Far from this case\textemdash particularly within the lossy regime\textemdash it is unclear if the redundancy stems from local classical knowledge on the pointer states of the qubit, since the plateau falls well below this bound. 
\begin{figure}[!t] 
	\centering
	\includegraphics[width=.48\textwidth]{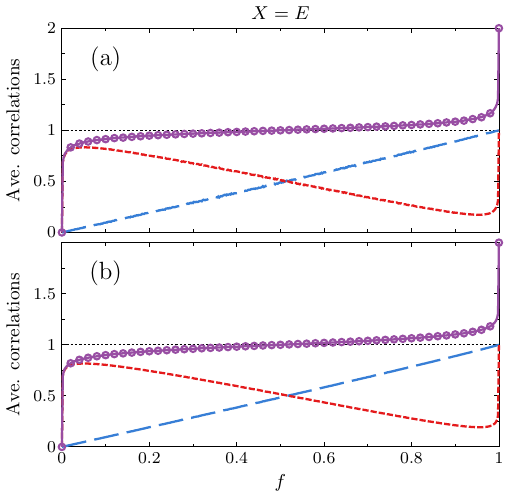}
	\caption{\label{AvecorrR}Total partial information (violet solid curve), partial accessible information (blue long-dashed curve), and partial quantum discord (red dashed curve) between the qubit and subenvironments [see Eqs. (\ref{eq:QMIsplit})-(\ref{eq:discord})], plotted over $S(\rho_S)$ at time $\Omega_0t=50$, with $\Gamma=10^{-3}\Gamma_+$ ($\Delta=0$). Snapshots are shown for parameters (a) $\Gamma_+=\Omega_0$ and (b) $\Gamma_+=10\Omega_0$. The sum of the classical and quantum correlations (violet open points) are also shown, indicating the validity of Eq. (\ref{eq:avecorr}).}
\end{figure}

\subsection{Accessible information and quantum discord}\label{Sec4b}

We shed light on the above discussion by considering the following definition of the QMI \cite{Henderson2001,OllivierDiscord2002}:
\begin{equation}\label{eq:QMIsplit}
	I(\rho_{SX_f}) = J(\rho_{SX_f}) + \bar{\delta}(\rho_{SX_f}),
\end{equation} 
where 
\begin{align}
	J(\rho_{SX_f}) &=\max_{\{M^{X_f}_j\}}\left[S(\rho_S)-S(\rho_S|\{M^{X_f}_j\})\right],\label{eq:Holevo} \\
	\bar{\delta}(\rho_{SX_f}) &=\min_{\{M^{X_f}_j\}}\left[S(\rho_{X_f})-S(\rho_{SX_f})+S(\rho_S|\{M^{X_f}_j\})\right].\label{eq:discord} 
\end{align}
The quantity $J(\rho_{SX_f})$ defines the upper limit of the Holevo bound \cite{Holevo1973}\textemdash the accessible information\textemdash which provides the maximum classical data that can be transmitted through the quantum channel. Accordingly, the conditional entropy $S(\rho_S|\{M^{X_f}_j\})$ of the bipartite system is written as
\begin{equation}\label{eq:condentropy}
	S(\rho_S|\{M^{X_f}_j\})= \sum_jp_jS\Big(\rho_{S|M^{X_f}_j}\Big), 
\end{equation}
which expresses the lack of knowledge in determining $\rho_S$ when $\rho_{X_f}$ is known. A measurement on the subsystem $X_f$ is formulated in terms of the projectors $M^{X_f}_j$, where the post-measurement state of the qubit is 
\begin{equation}
	\rho_{S|M^{X_f}_j}=\frac{1}{p_j}\text{Tr}_{X_f}\left[M^{X_f}_j\rho_{SX_f}M^{X_f}_j\right],
\end{equation}
with an outcome $j$ obtained with probability $p_j=\text{Tr}_{S,X_f}\left[M^{X_f}_j\rho_{SX_f}\right]$. The second quantity $\bar{\delta}(\rho_{SX_f})$ defines a general measure of quantum correlations between the two subsystems, known as the quantum discord \cite{Modi2012}. Note the bar is used to distinguish the discord from the information deficit $\delta$. \\
\indent It is emphasised that the accessible information and discord are optimised through a choice of positive operator-valued measure (POVM) $\{M^{X_f}_j\}$. Our motivation for minimizing the quantum discord stems from wanting to examine the correlations of the state least disturbed by measurement. Here, the measurement is formulated by mapping the relevant system fragment to an effective two-qubit state, as was done with Eqs. (\ref{eq:col_qubit_E}) and (\ref{eq:col_qubit_P}). The POVM $\{M^{X_f}_j\}$ then makes up a set of orthogonal projectors from the qubit states of $X_f$ (see appendix \ref{appenB}). \\
\indent Initially we compute Eqs. (\ref{eq:Holevo}-\ref{eq:discord}) for the full system-environment ($f=1$) of case (i) and find that the QMI is always shared equally between classical and quantum correlations when $I(\rho_{SE})>0$. This intuitively follows since the information encoded by classical data is limited to $S(\rho_S)$. The remaining information out of $S+E$ then has to make up the discord in equal amount assuming the state is pure, based on the global entanglement of the system and bath. Alternatively, for fractional states ($f<1$) we find a more interesting albeit complicated interplay between classical and quantum correlations. The quantities of interest here are the (averaged) partial accessible information $\langle J(\rho_{SX_f})\rangle$ and partial quantum discord $\langle \bar{\delta}(\rho_{SX_f})\rangle$, which from Eq. (\ref{eq:QMIsplit}), fulfil the relation 
 \begin{figure}[!t] 
 	\centering
	\includegraphics[width=.48\textwidth]{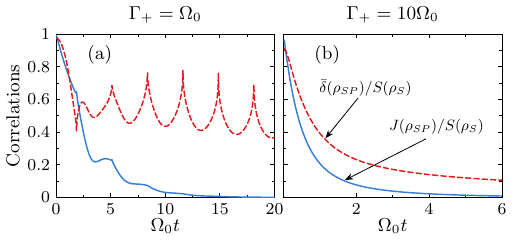}\\
	\includegraphics[width=.48\textwidth]{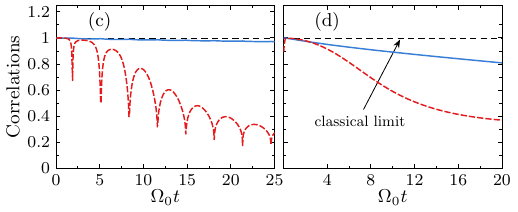}
	\caption{\label{AD}Dynamics of the accessible information (blue solid curve) and quantum discord (red dashed curve) between qubit and pseudomodes ($X=P$), taken from Eqs. (\ref{eq:Holevo}) and (\ref{eq:discord}) with $\Delta=0$ (solutions in appendix \ref{appenB}). Panels in the lefthand column are plotted for $\Gamma_+=\Omega_0$, while those in the right-hand column are plotted for $\Gamma_+=10\Omega_0$. (a), (b) $\Gamma=0.4\Gamma_+$: classical correlations decay on a fast time scale and quickly approach zero. (c), (d) $\Gamma=10^{-3}\Gamma_+$: classical correlations decay much slower with a large separation of time scales.}
\end{figure}
\begin{equation}\label{eq:avecorr}
	\langle I(f)\rangle=\langle J(\rho_{SX_f})\rangle+\langle\bar{\delta}(\rho_{SX_f})\rangle.
\end{equation}
In Fig. \ref{AvecorrR} we show plots of the average correlations for case (i). The most striking feature is the sharp rise in partial quantum discord around small fraction sizes. As quantum correlations generally decline in value for larger fractions, the accessible information grows linearly and as such is characteristic of non-redundant classical information\textemdash i.e., its partial information plot does not have a flat plateau shape. Note also that the distribution of classical correlations between different arrangements of fragments ($\langle J(\rho_{SX_f})\rangle$ vs. $f$) is essentially static over time and independent of system-environment coupling strength. The discord, which takes large values in the majority of fractions, therefore indicates a clear disturbance to the overall state from performing local measurements on the pseudomodes. This behavior reveals the interaction does \textit{not} produce a class of states exhibiting complete Darwinism.  \\
\begin{figure}[!t] 
	\centering
	\includegraphics[width=.48\textwidth]{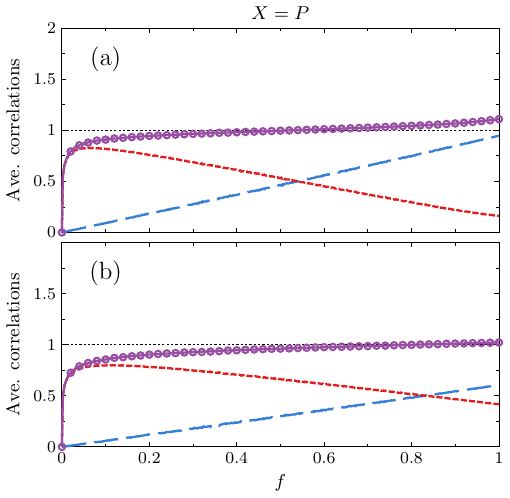}
	\caption{\label{AvecorrP} Total partial information (violet solid curve), partial accessible information (blue long-dashed curve), and partial discord (red dashed curve) between the qubit and pseudomodes plotted over $S(\rho_S)$ at time $\Omega_0t=50$, where $\Gamma=10^{-3}\Gamma_+$. (a), (b) Same parameters as those in Fig. \ref{AvecorrR}, i.e., (a) $\Gamma_+=\Omega_0$ and (b) $\Gamma_+=10\Omega_0$. Open points show the sum of the averaged classical correlations and discord, as in Eq. (\ref{eq:avecorr}).}
\end{figure}
 \indent Now turning our attention to case (ii), we address the dynamical behavior of the correlations with respect to the full fraction of pseudomodes, displayed in Fig. \ref{AD}. Let us start by considering the regime $\Gamma\ll\Gamma_+$. Remarkably, when the dynamics are non-Markovian, Eq. (\ref{eq:Holevo}) stays close to its maximum value over the course of the interaction, and hence the classical correlations are robust to the noise influence of the Markovian environment. This is also true but to a lesser extent in the case of Markovian dynamics\textemdash we recall that the effect of noise is more substantial with a higher rate of increase in vacuum population, which here increases the damping rate of the classical information by a factor proportional to $\Gamma_+/\Omega_0$ (e.g., a roughly ten times larger gradient between $\Gamma_+=\Omega_0$ and $10\Omega_0$). In contrast, the quantum discord begins to decay at a faster rate. At longer times it can be seen that the quantum correlations become better protected against decoherence when the discord decreases more slowly. We examine the limiting case of this behavior in Sec. \ref{2C}. Our current observation is that the dynamical behavior of the classical correlations is qualitatively the same for both weak and strong coupling. The only particular difference is the presence of memory effects in the latter. \\
\indent In the regime $\Gamma\approx\Gamma_+$ a somewhat opposite effect occurs with respect to the full qubit-pseudomode information. In this instance, classical correlations disappear asymptotically in time so that eventually the quantum discord makes up all of the QMI. As almost no classical information is present even within the full memory region of the environment, we find a case where quantum information is in fact redundant. This is confirmed in Fig. \ref{AD}(b) where the average correlation content of a fragment is shown with varying $f$. Here, $\langle\bar{\delta}(\rho_{SP_f})\rangle\approx \langle I(f)\rangle$ and $\langle J(\rho_{SP_f})\rangle\approx 0$ regardless of the size of the fragment. Moreover, Fig. \ref{AvecorrP}(a) shows these same quantities plotted for a large separation of time scales. We see that the quantum and classical correlations mimic those in Fig. \ref{AvecorrR} for case (i), though without a sudden increase in the discord at large $f$ since the maximum available information via measurement is limited when the state is mixed\textemdash even for the eigenstates of a global observable which coincide with the eigenbasis of $\rho_{SP}$ (i.e., partial information plots are not anti-symmetric). \\
\indent Overall, we conclude that the emergence of a classical plateau\textemdash in the partial information plots of either case (i) or (ii)\textemdash does not guarantee that classical information is redundant. This reveals significant underlying differences between the partial information plots presented in Figs. \ref{PIP(i)}\textemdash\ref{PIP(ii)} and those found in Refs. \cite{Paz2009,Zwolak2013}. For example, with an Ohmic environment interacting with a quantum Brownian oscillator, the quantum correlations (entanglement) are found to be suppressed for all but very large fraction sizes \cite{Paz2009}, meaning that the partial information plots alone reveal the presence of redundant classical correlations in the system. Here, we find that the same is not a sufficient condition for the redundancy of classical information, against what we originally interpreted as ``successful'' Darwinism in Sec. \ref{Sec4a}. A similar point regarding the nonunique association of the classical plateau to purely classically correlated states has been made in Ref. \cite{Horodecki2015}. 

\subsection{Maximization of classical correlations}\label{2C}

\begin{figure}[!t]
	\centering	
	\includegraphics[width=.46\textwidth]{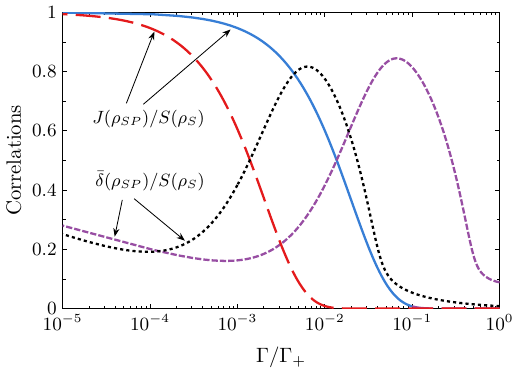}
	\caption{\label{CC_PM} Full accessible information [Eq. (\ref{eq:Holevo})] and quantum discord [Eq. (\ref{eq:discord})] for case (ii), plotted as a function of $\Gamma$ at time $\Omega_0t=50$ ($\Delta=0$). The blue solid and violet dashed curves are for $\Gamma_+=\Omega_0$, while the red long-dashed and black dotted curves are for $\Gamma_+=10\Omega_0$.}
\end{figure}

So far, in studying case (ii) we have found that for a Lorentzian (\ref{eq:SFplus}) with a highly peaked internal structure, $\Gamma\ll\Gamma_+$, the accessible information is non-redundant and significantly delocalized across the environment. As a corollary to the results of Sec. \ref{2B}, we examine the circumstances under which the classical correlations are maximised against the quantum discord in ``global'' correlations, e.g., for $f=1$.  \\
\indent Whether classical or quantum correlations are predominant has been shown not to depend on the presence of memory effects in the dynamics. This suggests it depends only on the degree of separation of the timescales. In fact, numerical evidence shown in Fig. \ref{CC_PM} reveals that decreasing the ratio $\Gamma/\Gamma_+$ further slows down the decay of classical correlations compared to the plots shown in the bottom row of Fig. \ref{AD}. If we observe the behavior $J(\rho_{SP})$ at a fixed time, we see it grows larger by decreasing the value of $\Gamma$. This behavior lies in contrast to the quantum correlations which tend to fall off more quickly, but can still make up a larger proportion of the total correlations given the QMI also dissipates more slowly. We postulate that in the idealised limit of Eq. (\ref{eq:largesep}), i.e., with a large separation of time scales,
\begin{equation}\label{eq:maxred}
		\Gamma t\longrightarrow 0, \qquad  \Gamma_+t\gg 1, \qquad \Omega_0/\Gamma_+=\text{fixed},
\end{equation}
the accessible information converges towards its maximum, thereby revealing that the full memory region of the environment acquires (almost all) classical data on the qubit state. Notice the limit $\Gamma_+t\rightarrow\infty$ is avoided, as here the qubit would approach the ground state with zero entropy, resulting in all correlations being lost. Let us now assume Eq. (\ref{eq:maxred}) holds. As $t$ increases further, what we expect is for the discord to make up an increasingly smaller proportion of the QMI. Once $J(\rho_{SP})\gg\bar{\delta}(\rho_{SP})$ in the very long time limit, the state $\rho_{SP}$ then shows robustness under non-selective measurements $\{M^{P}_j\}$ ($f=1$). Writing this in terms of a local operation on $P$, $(\Lambda_P\otimes \mathbbm{1}_S)\rho_{SP}$, we should have
\begin{align}
	\left(\Lambda_{P}\otimes \mathbbm{1}_S\right)\rho_{SP} &= \sum_j\left(\mathbbm{1}_S\otimes M^{P_f}_j\right)\rho_{SP}\left(M^{P_f}_j\otimes \mathbbm{1}_S\right)\nonumber\\
	&\approx \rho_{SP}.
\end{align}
Of course, the finite nature of the quantum correlations means our results do not provide an example of complete einselection \cite{OllivierDiscord2002}, yet it can be appreciated that the state attains its most classical-like form when the dynamics fulfill Eq. (\ref{eq:maxred}). It is also interesting to note how the slow loss of classical correlations occurs in line with the slow decay in the pseudomode population, which as we recall from Sec. \ref{2B} occurs past the crossover in dynamics at times $t\sim O(1/\Gamma_+)$.

\subsection{Non-Markovianity}\label{Sec4d}

Now we consider the relation between the non-Markovianity of the qubit dynamics and the redundancy measure (\ref{eq:red}). Our model provides a good foundation to investigate this connection as there is clear delineation of memory effects in terms of information backflow to the open quantum system. \\
\indent We illustrate the concepts surrounding non-Markovianity in this model by starting with the following definition. Let us state that if the family of dynamical maps $\Phi(t,0)$ governing the evolution
\begin{equation}\label{eq:dynmap}
	\rho_S(0)\mapsto\rho_S(t) = \Phi(t,0)\rho_S(0),\quad t\geq 0
\end{equation}
is divisible into two completely positive and trace preserving maps (CPTP), that is,
\begin{equation}\label{eq:divisdynmap}
	\Phi(t_2,0)=\Phi(t_2,t_1)\Phi(t_1,0),\quad\forall t_2\geq t_1\geq0,
\end{equation}
then $\rho_S$ undergoes a Markovian evolution. Here the notion of divisibility is enough to distinguish between what is considered Markovian and non-Markovian behavior \cite{Wolf2008,*Rivas2014,*Luo2012,*Rivas2010,*DeVega2017}. However, it should be stressed that CPTP maps associated with time-dependent Markov processes are categorically different from those which form a dynamical semigroup \cite{BreuerTOQS2002}. In a similar way, the situation where a given amount of information leaks back to the open system has been shown to be intimately related to the non-Markovian properties of the quantum channel. The Breuer-Laine-Piilo (BLP) measure of non-Markovianity \cite{Breuer2009}, as an example, is rooted in the particular interpretation of shared information as the distinguishability of a pair of input states $\{\rho^1_S,\rho^2_S\}$ to the channel. Their distinguishability is characterized by the trace distance $D(\rho^1_S,\rho^2_S)$, given by
\begin{equation}\label{eq:tracedis}
	D(\rho^1_S,\rho^2_S) =\frac{1}{2}\text{Tr}\left|\rho^1_S(t)-\rho^2_S(t)\right|,
\end{equation}
where $|A|=\sqrt{A^{\dagger}A}$. Under a divisible map (\ref{eq:divisdynmap}) the rate of change in the trace distance over any time interval is always negative. One can then show that the distinguishability of the pair of states is a monotonically decreasing function in time\textemdash overall, corresponding to a continual loss of information from $S$ to $E$ \cite{Breuer2016}. Variation from this behavior, i.e., when the trace distance temporally increases, indicates that the process is nondivisible through reverse flow of information back to the open system. In terms of the rate of change of the trace distance, 
\begin{equation}
	\sigma(\rho^{1,2}_S(0),t)=\dot{D}(\rho^1_S,\rho^2_S),
\end{equation}
the quantum process is non-Markovian if and only if $\sigma(t)>0$ at some point during the evolution of the density matrices $\rho^{1,2}_S(t)=\Phi(t,0)\rho^{1,2}(0)$. \\ 
\indent Concerning our model, the trace distance can always be expressed analytically as
\begin{equation}\label{eq:tracedis2}
	D(\rho^1_S,\rho^2_S) = |G(t)|\sqrt{|G(t)|^2a^2+|b|^2},
\end{equation}
where $a=\bra{e}\rho^1_S(0)\ket{e} - \bra{e}\rho^2_S(0)\ket{e}$ and $b=\bra{e}\rho^1_S(0)\ket{g} -\bra{e}\rho^2_S(0)\ket{g}$. By taking the time derivative of the above \cite{Laine2010},
\begin{figure}[!t] 
	\centering
	\includegraphics[width=.48\textwidth]{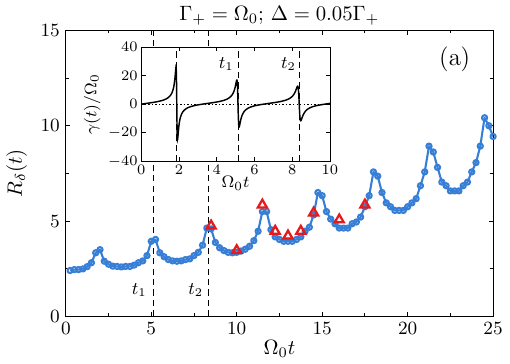}\\
	\includegraphics[width=.48\textwidth]{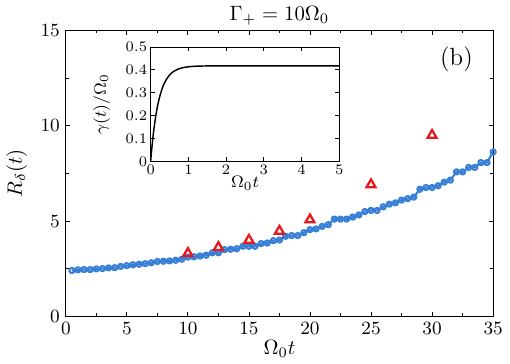}
	\caption{\label{Fig8} Redundancy $R_{\delta}(t)$, Eq. (\ref{eq:red}), plotted as a function of time for cases (i) (red triangles) and (ii) (blue marked curve) with $\delta=0.15$. Insets: Decay rate $\gamma(t)$ from Eq. (\ref{eq:timedepdecayrate}) for each set of parameters. (a) $\Gamma_+=\Omega_0$ and $\Delta=0.05\Gamma_+$: the dashed lines indicate times at which the redundancy peaks. (b): $\Gamma_+=10\Omega_0$ and $\Delta=0$: in the weak-coupling limit, redundancy monotonically increases in time.}
\end{figure}
\begin{equation}\label{eq:sigma}
	\sigma(t,\rho^{1,2}_S(0)) = \frac{2|G(t)|^2a^2+|b|^2}{\sqrt{|G(t)|^2a^2+|b|^2}}|\dot{G}(t)|, 
\end{equation}
we see that $\sigma(t)$ is only positive if $|\dot{G}(t)|>0$ at any time $t$, which here sets the condition for nondivisibility and thus non-Markovianity from the BLP measure \cite{Breuer2012}. Hence information flows back to the qubit at times when there are temporary revivals in its population. Since the master equation (\ref{eq:SMe}) also has a time-local structure, we can use Eqs. (\ref{eq:gm(t)}) and (\ref{eq:sigma}) to prove that
\begin{equation}
	\sigma(t,\rho^{1,2}_S(0))>0\longleftrightarrow\gamma(t)<0,\quad\forall t\geq0.
\end{equation}
It is therefore reasonable to suspect that memory effects will appear in the partial information plots at such times when the dynamics are non-Markovian\textemdash specifically, in instances when the decay rate (\ref{eq:timedepdecayrate}) becomes negative. How the redundancy behaves with respect to changes in the trace distance (\ref{eq:tracedis2}) and $\gamma(t)$ is precisely the connection we aim to make with our results. \\
\indent In Fig. \ref{Fig8}, we show various values of the redundancy computed in both the strong- and weak-coupling limits. Note for $\Gamma_+=\Omega_0$ the detuning is taken to be non-zero, for the reason that the map $\Phi(t,0)$ is noninvertible for $\Delta=0$, and so no strict definition of divisibility exists in this case \cite{Breuer2016}. Here, we are not necessarily interested in the exact numerical value of the redundancy, but rather its dynamical behavior with respect to the decay rate. Since, from Eq. (\ref{eq:gm(t)}), we have established that the qubit receives back population directly from the depletion of the pseudomodes when the rate in the master equation is negative, it is plausible to think memory effects will also influence correlations between $S$ and $P$. Therefore, we have additionally computed the redundancy for case (ii)\textemdash using a large separation of time scales\textemdash to compare with the those of case (i). \\
\indent In the strong-coupling regime, the key indication from our results is that the redundancy has peaks and troughs exactly in line with the decay rate. Let us first consider case (ii) at time intervals during which $\gamma(t)>0$. Here, the redundancy is seen to increase up until the point at which the decay rate is discontinuous and subsequently becomes negative. It is noticed that the plateau grows in length as the open system monotonously loses information into the environment. Then, as $\gamma(t)$ begins to grow from negative values, that is, when information flows back to the qubit, the redundancy plateau is suppressed considerably before increasing again at times when the decay rate becomes positive. The same type of behavior can also be seen to occur from the perspective of case (i), where memory effects are visible during times when the plateau emerges. \\
\indent Alternatively, in the weak-coupling regime the plateau only continuously grows in length while the qubit undergoes a Markovian evolution [in both cases (i) and (ii)]. This firmly suggests that the nonmonotonicity of the redundancy captures the non-Markovian dynamics of the qubit\textemdash which, in this context, is clearly based on the same effective behavior in the trace distance (\ref{eq:tracedis2}). We also point out that the connection between quantum Darwinism and non-Markovianity has also been studied recently in Refs. \cite{Giorgi2015,Galve2016}. The authors find a similar effect, where information backflow to the open system translates into poor Darwinism, i.e., a worsening of the plateau at these times. \\
\indent Furthermore, the fact that the redundancy of cases (i) and (ii) shows similar dynamical behavior suggests that the rollback of the plateau occurs specifically because of information backflow \textit{from} the pseudomodes (memory) \textit{to} the qubit. Indeed, consider the example of an initially excited qubit for the same parameters as Fig. \ref{Fig8}(a). What we find is that at times when $|G(t)|^2$ starts to increase there is an accompanied increase in QMI of the qubit and environment from its minimum zero value. This indicates that correlations previously removed by dissipation re-develop because of information flow back at times when the redundancy decreases. Now, within the regime $\Gamma\ll\Gamma_+$ (shown in Fig. \ref{Fig8}), we observe that the dynamics of QMI between the qubit and memory $I(\rho_{SP})$ faithfully coincides with that of $I(\rho_{SE})$. Thus we can associate the same re-correlation effect with revivals in the qubit population, which, from Eq. (\ref{eq:gm(t)}), simultaneously comes from losses in the pseudomodes. In this sense the energy/information received back by the qubit from the pseudomodes can provide the physical mechanism for the drop in the classical plateau.

\section{Summary and discussion}\label{Sec5}

In summary, we have provided a detailed investigation into emergent features of quantum Darwinism by applying the framework to a qubit interacting with many subenvironments of bosons. The basis of our paper derives from the idea that the original environment maps to a bipartite structure containing a memory and non-memory part. From this we have examined how information is encoded into fractions of the environment in two separate cases: one where we constructed random fragments out of subenvironments [case (i)], and the other where we constructed random fragments of independent parts of the full memory [case (ii)]. Our main effort has been to recognize whether the emergence of redundant information occurs, and, if so, where into the environment such information is proliferated. By considering different dynamical regimes we identify instances in cases (i) and (ii) where redundant information forms close to the classical bound, implying ``successful'' Darwinism. 

Despite the formation of the classical plateau, which usually is considered as a hallmark of successful quantum Darwinism, our results demonstrate a scenario where classical information is precisely non-redundant. Consequently, we found the quantum discord\textemdash taken from the partial information\textemdash to obtain relatively large values in small fractions of pseudomodes and/or subenvironments, realizing the highly non-classical nature of a typical system-fragment state based on the fact that it is disturbed significantly under local (projective) measurements. In parallel we have analyzed the dynamics of the classical and quantum correlations between the qubit and memory region. In both cases\textemdash either when considering the partial correlations from fractions of the environment or full collection of pseudomodes\textemdash we have found qualitatively similar behavior in the results across both the strong- and weak-coupling regimes. Substantial differences are only introduced through relatively varying the decay rates of the pseudomode population. For example, in the lossy regime, that is, where correlations and population are significantly damped in time, the QMI of the qubit-pseudomode system shows asymptotically decaying classical correlations with prevalent discord. A regime where the pseudomodes maximize their classical correlations over the course of the dynamics has also been identified. 

Finally, we have sought to cement the connection between the emergence of redundancy in the quantum Darwinism framework and non-Markovianity. Memory effects are characterized by the backflow of information (and population), which in turn reflects the nondivisibility of the dynamical map. We show that the redundancy plateau in the partial information plots (see Fig. \ref{Fig8}) is suppressed at times when information flows from the pseudomodes to the qubit. Remarkably, the redundancy acts as a witness to non-Markovian behavior in directly the same way as the trace distance does in the BLP measure.  

Looking ahead, our results support the possibility of developing a novel quantifier of non-Markovianity from the perspective of information redundancy in line with Ref. \cite{Galve2016}. While the redundancy $R_{\delta}(t)$ does not strictly detect memory effects on the basis of divisibility but instead from information flow between the system and environment, these two concepts have been shown to be closely connected \cite{Haseli2014,Bylicka2014} through measures of information (e.g., the QMI) similar to those employed in the current paper. This could lead to comparisons between related measures beyond the current model. Alternatively, the methods employed here could be applied to practical models of cavity QED systems to test for similar effects encountered in this paper. \\
\indent\textit{Note added in proof}. Recently, we became aware of \cite{Lampo2017}, which examines the effect of non-Markovianity on the emergence of classicality (and hence quantum Darwinism) in a similar way as Refs. \cite{Giorgi2015,Galve2016}.

\section{Acknowledgements}

This paper has been supported by the UK Engineering and Physical Sciences Research Council (Grant No. EP/L505109/1). 

\appendix

\section{Exact solutions for the state coefficients}\label{appenA}

For a sufficiently large number of subenvironments $\#E$ the weights $w_k$ may be replaced across the interval $d\xi_k$ by 
\begin{equation}
	w_k = W(\xi_k)d\xi_k,
\end{equation}
with $W(\xi)$ modeling  the underlying pseudomode distribution in Eq. (\ref{eq:SF}). Imposing the continuum limit $\#E\rightarrow\infty$, the memory kernel is written as   
\begin{equation}\label{eq:seckern}
	f(t) = \Omega^2_0\int^{\infty}_{-\infty}d\xi\,W(\xi)\,\text{exp}[-i(\xi-\omega_0)t-\Gamma t/2].
\end{equation} 
Note that $W(\xi)$ is in principle arbitrary, apart from having to fulfill the normalization condition $\int^{\infty}_{-\infty}d\xi W(\xi)=1$. For simplicity, we take 
\begin{equation}\label{eq:PMdis}
	W(\xi) = \frac{1}{\pi}\frac{\Gamma_W/2}{(\omega_0-\Delta-\xi)^2+(\Gamma_W/2)^2}, 
\end{equation} 
with the parameters being defined in the main text. The function contains a single pole in the lower complex $\xi$-plane such that Eq. (\ref{eq:seckern}) can be evaluated using the residue theorem. The resulting equation for $G(t)$ is solved using Laplace transforms, where
\begin{equation}\label{eq:qsol}
	G(t) = e^{(i\Delta/2-\Gamma_+/4)t}\left[\cos\left(\frac{\Omega t}{2}\right) - \frac{(i\Delta - \Gamma_+/2)}{\Omega}\sin\left(\frac{\Omega t}{2}\right)\right],
\end{equation}   
and $\Omega=\sqrt{4\Omega^2_0-(i\Delta-\Gamma_+/2)^2}$. For the sake of completeness, we note the same can be derived by noting that the structure is given by the convolution $D(\omega)=(W\ast L)(\omega)$, where
\begin{equation}
	(W\ast L)(\omega) = \int^{\infty}_{-\infty}d\xi\, W(\xi)L(\omega-\xi),
\end{equation}
and $L(\omega) = \Gamma/[\omega^2 + (\Gamma/2)^2]$. The result is simply another Lorentzian of width $\Gamma_+$, as given in Eq. (\ref{eq:SFplus}). \\
\indent The coefficients $c_{k,\lambda}(t)$ are found by substituting Eq. (\ref{eq:qsol}) into the Schr\"{o}dinger equation, while $b_k(t)$ are obtained through Eq. (\ref{eq:pseud}). Their solutions are given by 
\begin{widetext}
\begin{align}
\begin{split}\label{eq:b_k}
	b_k(t) &= -\frac{4\Omega_k\,c_e(0)}{\left(i(2\Delta_k+\Delta)+\frac{\Gamma-\Gamma_W}{2}\right)^2 + \Omega^2}\Bigg\{\bigg(\Delta_k+\Delta+i\frac{\Gamma_W}{2}\bigg)e^{-\Gamma t/2}\\
	&\qquad - e^{i\Delta_kt}e^{(i\Delta/2-\Gamma_+/4)t}\Bigg[\left(\Delta_k+\Delta+i\frac{\Gamma_W}{2}\right)\cos\left(\frac{\Omega t}{2}\right) \\
	&\qquad - \left(\left(\frac{i\Delta - \Gamma_+/2}{\Omega}\right)\left(\Delta_k+\Delta/2-i\frac{\Gamma-\Gamma_W}{4}\right)+ i\frac{\Omega}{2}\right)\sin\left(
	\frac{\Omega t}{2}\right)\Bigg]\Bigg\}, 
\end{split}
\\
\begin{split}
	c_{k,\lambda}(t) &= -\frac{4g_{k,\lambda}\,c_e(0)}{(i(2\delta_{\lambda}+\Delta)-\Gamma_+/2)^2+\Omega^2}\Bigg\{\left(\delta_{\lambda}+\Delta+i\frac{\Gamma_+}{2}\right)\\
	&\qquad -e^{i\delta_{\lambda}t}e^{(i\Delta/2-\Gamma_+/4)t}\Bigg[\left(\delta_{\lambda}+\Delta+i\frac{\Gamma_+}{2}\right)\cos\left(\frac{\Omega t}{2}\right) \\
	 &\qquad - \left(\left(\frac{i\Delta-\Gamma_+/2}{\Omega}\right)(\delta_{\lambda}+\Delta/2+i\Gamma_+/4) + i\frac{\Omega}{2}\right)\sin\left(\frac{\Omega t}{2}\right)\Bigg]\Bigg\},
\end{split}
\end{align}
\end{widetext}
where $\delta_{\lambda}=\omega_{\lambda}-\omega_0$. To compute the relevant von Neumann entropies we trace over the degrees of freedom excluded from $X_f$ using the two-qubit representation of the density matrix $\rho_{SX_f}$. For example, a single realization of $\rho_{SX_f}$ is composed by eliminating $(\#E-m)$ randomly selected (i) subenvironments from the pure state (\ref{eq:state}) or (ii) pseudomodes from the mixed state solution of Eq. (\ref{eq:PMme}). The latter is given by
\begin{equation}\label{eq:PMdm}
	\rho_{SP} = \Pi_p(t)\ket{g}\bra{g}\otimes(\ket{0}\bra{0})_P + (\ket{\widetilde{\psi}(t)}\bra{\widetilde{\psi}(t)})_{SP}, 
\end{equation}
where $\ket{0}_P$ denotes the collective vacuum of the pseudomodes. In addition, the the un-normalized state vector $\ket{\widetilde{\psi}(t)}$ reads 
\begin{equation}
	\ket{\widetilde{\psi}(t)} = c_g\ket{g,0}+c_e(t)\ket{e,0}+\sum_{k}b_k(t)\ket{g,1_k},
\end{equation}
where $\ket{g,1_k}$ indicates a single excitation in the $k$ pseudomode. The pseudomode vacuum population is given by
\begin{equation}
	\Pi_p(t) = \Gamma\int^t_0ds\sum_k|b_k(s)|^2.
\end{equation}

\section{Accessible information and quantum discord}\label{appenB}

To gain a clear understanding on the distinction between classical and quantum information, we use the Holevo quantity in Eq. (\ref{eq:Holevo}) to gauge the information accessible via measurements on $X_f$. This information is limited by the type of measurement used. Because the QMI is invariant to how $J(\rho_{SX_f})$ and $\bar{\delta}(\rho_{SX_f})$ are assigned, the quantum discord, in turn, is required to be minimized over all measurement bases $\{M^X_j\}$ to avoid erroneous results. The POVM that fulfils this condition has been shown by Datta to be formulated using rank-1 projectors \cite{ADatta2008}. \\
\indent In order to realize the measurement in terms of such projectors, we make use of the fact that both the subenvironments and pseudomodes\textemdash or fractions thereof\textemdash can collectively be mapped to a single qubit. As stated in the main text, the ground state of the qubit is defined from the vacuum of $X_f$: $\ket{\widetilde{0}}_{X_f}=\ket{\{0\}}_{X_f}$, while the excited state $\ket{\widetilde{1}}_{X_f}$ is formed through Eqs. (\ref{eq:col_qubit_E}) and (\ref{eq:col_qubit_P}). Let us write the complete set of local orthogonal projectors in terms of the qubit states:
\begin{align}
		M^{X_f}_1 &= \frac{1}{2}\left(\mathbbm{1}_{X_f} + {\bm r}\cdot{\bm \sigma}\right),\\
		M^{X_f}_2 &= \frac{1}{2}\left(\mathbbm{1}_{X_f} - {\bm r}\cdot{\bm \sigma}\right), 
\end{align}
where ${\bm r}=\left(\sin\theta\cos\phi,\sin\theta\sin\phi, \cos\theta\right)^T$ is the Bloch vector, and ${\bm \sigma}=\left(\sigma_x,\sigma_y,\sigma_z\right)^T$ contains the Pauli operators constructed from the basis $\{\ket{\widetilde{0}}_{X_f},\ket{\widetilde{1}}_{X_f}\}$, along with the identity $\mathbbm{1}_{X_f}$. The accessible information and discord are then extremized with respect to the free choice of angles $\theta\in[0,\pi)$ and $\phi\in[0,2\pi)$. \\
\indent Remembering that the conditional state is $\rho_{S|M^{X_f}_j}=\text{Tr}_{X_f}[(\mathbbm{1}_S\otimes M^{X_f}_j)\rho_{SX_f}(M^{X_f}_j\otimes\mathbbm{1}_S)]/p_j$ ($j=1,2$), for each measurement one obtains 
\begin{align}\label{eq:condden}
	\rho_{S|M^{X_f}_j} &= \frac{1}{p_j}\bigg(A_j(\theta)\ket{g}\bra{g} + C_j(\theta)\ket{e}\bra{e}\nonumber\\
				     &+ B_j(\theta,\phi)\ket{e}\bra{g} + \text{h.c.}\bigg)
\end{align}
where 
\begin{equation}
\begin{split}
	A_j(\theta) 		&= \frac{1}{2}\Big\{\Pi_{X_f}(t)+\eta^2_{X_f}(t)\\
					&\quad+ (-1)^{j-1}\cos\theta\Big[\eta^2_{X_f}(t)-\Pi_{X_f}(t)\Big]\Big\}, \\
	B_j(\theta,\phi) 	&= \frac{1}{2}(-1)^{j-1}\sin\theta\,e^{-i\phi}\eta_{X_f}(t)c_e(t), \\
	C_j(\theta) 		&= \frac{1}{2}|c_e(t)|^2\Big(1+(-1)^j\cos\theta\Big), 
\end{split}
\end{equation} 
with probabilities $p_j = \frac{1}{2}\big\{1+(-1)^j\cos\theta\left[\langle\sigma^S_z\rangle+2\Pi_{X_f}(t)\right]\big\}$. The coefficients $\eta_{X_f}(t)$ and $\Pi_{X_f}(t)$ are given by 
\begin{equation}
	\eta_{X_f}(t) = 
	\begin{cases}
    		\sum_{k\in E_f,\lambda}|c_{k,\lambda}(t)|^2,&\text{if }X=E,\\
    		\sum_{k\in P_f}|b_k(t)|^2,& \text{if }\,X=P,
	\end{cases}
\end{equation}
and 
\begin{equation}
	\Pi_{X_f}(t) = 
	\begin{cases}
    		|c_g|^2+\sum_{k\in E_{1-f},\lambda}|c_{k,\lambda}(t)|^2,&\text{if }X=E,\\
    		|c_g|^2+\sum_{k\in P_{1-f}}|b_k(t)|^2+\Pi_p(t),& \text{if }\,X=P,
	\end{cases}
\end{equation}
where $k\in X_{1-f}$ denotes summation over objects \textit{not} in the fragment. \\
\indent Finally, by diagonalizing Eq. (\ref{eq:condden}) its eigenvalues are obtained and substituted into Eq. (\ref{eq:condentropy}) to evaluate the conditional entropy. We get 
\begin{equation}
	S(\rho_{S|\{M^{X_f}_j\}}) = -\sum_{i,j=1,2}p_j\lambda_{i,j}\,\text{ln}\,\lambda_{i,j},
\end{equation}
where
\begin{equation}
	\lambda_{i,j} =  \frac{A_j(\theta) + C_j(\theta) + (-1)^i\sqrt{\left[A_j(\theta)-C_j(\theta)\right]^2 + 4|B_j(\theta,\phi)|^2}}{2p_j}. 
\end{equation}
From the above expression it is easy to see that the conditional entropy is invariant with respect to $\phi$. 

\bibliographystyle{apsrev4-1} 

\bibliography{QD_references}

\end{document}